\documentclass[aps,prl,twocolumn,float]{revtex4}
\usepackage{amsmath,bm,epsfig}
\newcommand{\B}[1]{{\bm{#1}}}
\begin{document}

\title{Shear Transformation Zones: State Determined or Protocol Dependent?}
\author{Oleg Gendelman$^1$, Prabhat K. Jaiswal$^2$, Itamar Procaccia$^2$, Bhaskar Sen Gupta$^2$, and Jacques Zylberg$^2$}
\affiliation{$^1$ Faculty of Mechanical Engineering, Technion, Haifa 32000, Israel\\$^2$Weizmann Institute of Science,  Rehovot 76100, Israel}
\begin{abstract}
The concept of a Shear Transformation Zone (STZ) refers to a region in an amorphous solid that undergoes a plastic
event when the material is put under an external mechanical load. An important question that had accompanied the development
of the theory of plasticity in amorphous solids for many years now is whether an STZ is a {\em region} existing in the material (which can be predicted by analyzing the unloaded material), or is it an {\em event} that depends on the loading protocol (i.e., the event cannot be predicted without following the protocol itself). In this Letter we present strong evidence that the latter is the case. Infinitesimal changes of protocol result in macroscopically big jumps in the positions of plastic events, meaning that these can never be predicted from considering the unloaded material.
\end{abstract}
\maketitle
The origin of plastic responses to external mechanical loads in crystalline solids is understood: topological defects, and in particular dislocations, glide under the action of external stresses or strains, and this glide is irreversible, dissipating energy as it
is taking place \cite{MS73,C53,11Tol}. Of course, when the density of such defects increases, the situation becomes hairy, and proper theories are still
under active research. The fundamental mechanism of  plasticity in amorphous solids is, on the other hand, still not fully resolved.
In essence there are two schools of thought. The first considers plasticity resulting from the existence of some regions in the
material that are more sensitive to external load. These regions are referred to as Shear Transformation Zones (STZ) and their
introduction to rheological models of amorphous solids goes back to the work of Argon, Spaepen, and Langer \cite{Arg,80TS,98FL,07BLP}. The second school considers
plasticity as an instability of the amorphous solids \cite{98ML,04ML,KLLP10} resulting from a protocol of an increase in the external load. This instability can be understood by focusing on the Hessian matrix of the material (and see below for more details) with an eigenvalue that
goes to zero following a saddle-node bifurcation \cite{09LP,12DKP,11HKLP}. Both schools of thought agree that until the appearance of system spanning plastic events (shear bands) at high values of the external load, the plastic events that one is discussing are localized. In the instability way of thinking this is explained by the localization of the eigenfunction associated with the eigenvalue that is going to zero.

 The difference in thought is not only in choosing words to describe plasticity
in amorphous solids. If the STZ approach is valid, one should be able to predict, by a judicious analysis of the unstrained
system, where a plastic event is likely to take place. If indeed there are some regions that are more sensitive than others
to external loads, they should be identifiable and marked prior to exercising the external load. On the other hand, if the
protocol dependence of an instability is the right way of thinking, then one should be able to show that even minute changes
in protocol will result in a major change in the plastic event that may take place. Then it would be argued that it were not
possible to predict where plasticity should appear. The aim of the present Letter is to propose simple numerical simulations
that can decide between the two possibility, with the proposed result that the second way of thinking should prevail.

In our simulations we construct a 2-dimensional glass forming system in the usual way \cite{94KA}, selecting a binary mixture of $N$ particles, 50\% particles A and 50\% particles B, interacting via Lennard-Jones potentials. The difference between the particles is in the positions and the depths of the minima of the potentials; we choose the positions of the minima such that $\sigma_{AA}=1.17557$, $\sigma_{AB} =1.0$, and $\sigma_{BB}=0.618034$. The depths of the potentials are $\epsilon_{AA}=\epsilon_{BB}=0.5$ and $\epsilon_{AB}=1.0$. Below lengths
and energies are measured in units of $\sigma_{AB}$ and $\epsilon_{AB}$. The potential
is truncated at $r_{co}=2.5$ and goes smoothly to zero (with two derivatives). These parameters are
known to guarantee good glass formation and the avoidance of crystallization.

\begin{figure}
\includegraphics[scale = 0.25]{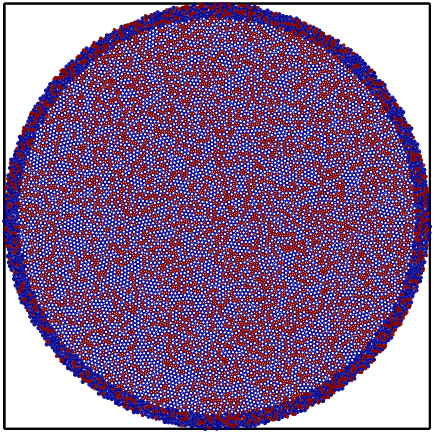}
\caption{The system with circular symmetry constructed as described in the text. In this example the total number of particles
is $N=20,000$.}
\label{sys1}
\end{figure}

The system is first equilibrated in a square box
of length $2R$ at a high temperature ($T=0.8$) with periodic boundary conditions. Secondly, the system is quenched to temperature  $T=0.001$ at constant volume
by molecular dynamics. Lastly, the system is energy minimized to $T=0$. At this point we build from the given
configuration a sub-system with circular symmetry using
the following protocol: we discard all the particles
outside a circle of radius R, fixing the positions of particles in an annulus (wall) of width $dR=2r_{co}=5.0$. An example of the resulting system with $N=20,000$ is shown in Fig. \ref{sys1}. Needless to say, once we fix the wall the periodic boundary conditions are lost.

\begin{figure}
\includegraphics[scale = 0.40]{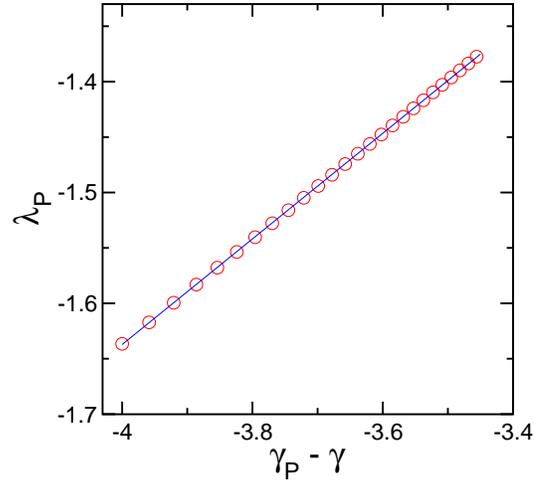}
\caption{A log-log plot of $\lambda_P$ vs. $\gamma_P-\gamma$. The measured slope is $0.48\pm 0.0004$.}
\label{lamvsgam}
\end{figure}

Having constructed a system with circular symmetry we can now follow its response to external loading. We load the system
athermally and quasi-statically, pulling along the $x$-axis and compressing along the $y$-axis such as to conserve the area.
Thus, the circular system that begins with $L_x =L_y=R$ deforms to an ellipse with principal axes $L_x\ne L_y$.
The affine step is area preserving, written as
\begin{eqnarray}
x' &=& x (1+ \delta \gamma)\ , \\
y' &=& \frac{y}{1+ \delta \gamma} \ .
\label{xy}
\end{eqnarray}
Note that in this affine steps also the wall particles are participating, to hold the system as desired. 
After every affine step of loading we annul the forces between the bulk particles (excluding the wall particles) using gradient energy minimization. The system
then undergoes a non-affine step that brings the system back to mechanical equilibrium.  This quasi-static loading is continued
as long as the system responses reversibly. The mechanical stability of the system is determined by  the Hessian matrix $\B H$:
\begin{equation}
H_{ij} \equiv \frac{\partial^2 U}{\partial \B r_i\partial \B r_j} \ ,
\end{equation}
where $U(\B r_1,\B r_2,\cdots \B r_N)$ is the total potential energy of the system as a function of
 the particle positions $\{\B r_i\}_{i=1}^N$. The Hessian matrix is real, symmetric, and positive definite as long as the system
 is mechanically stable, the first plastic event occurs when the lowest eigenvalue of $\B H$ approaches zero.
 It is well known that this happens via a saddle-node bifurcation, meaning that as a function of $\gamma=\sum{\delta \gamma}$
 there exists a value $\gamma=\gamma_P$ where the lowest eigenvalue $\lambda_P$ approaches zero via a square-root singularity
 \begin{equation}
 \lambda_P \sim \sqrt{\gamma_P-\gamma} \ .
 \end{equation}
 An example of this scaling law is presented in Fig. \ref{lamvsgam}.

\begin{figure}
\includegraphics[scale = 0.30]{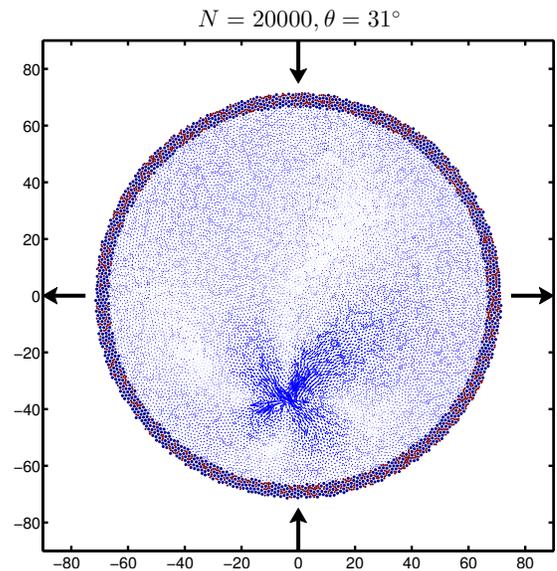}
\caption{The first plastic event that occurs as a result of choosing the $x$-axis to be in 31$^o$ with respect to the
horizontal direction of the original square box.  In this example $N=20,000$.}
\label{squeez1}
\end{figure}

\begin{figure}
\includegraphics[scale = 0.30]{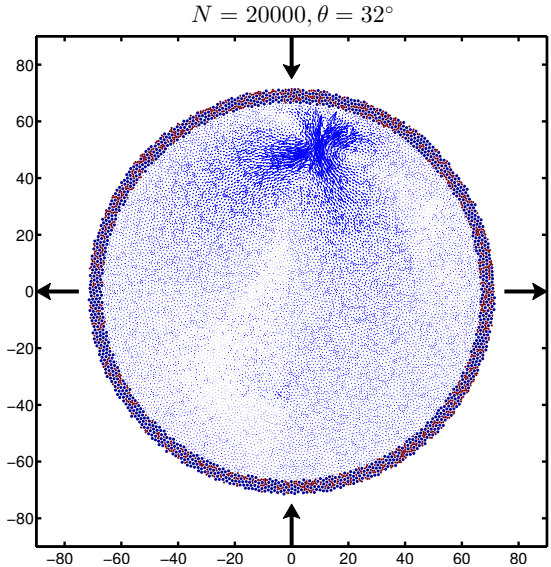}
\caption{Same as in Fig. \ref{squeez1} but with choosing the $x$-axis to be in 32$^o$ with respect to the
horizontal direction of the original square box.}
\label{squeez2}
\end{figure}

In the unloaded state all the eigenfunctions of the Hessian matrix which are associated with low lying eigenvalues are delocalized. Upon the approach of the lowest eigenvalue to zero, the associated wave-function $\Psi_P$ localizes on a typical quadrupolar structure which
is identical with the non-affine irreversible displacement associated with the plastic instability. An example of this phenomenon
is shown in Fig. \ref{squeez1} which is obtained by selecting the $x$-axis in Eq.~(\ref{xy}) to be at 31$^o$ with respect to the
horizontal direction of the original square box from which we constructed the circularly symmetric system. The plastic event is shown
as the quadrupolar displacement field near the bottom of the system. One can at will call it an STZ, but consider what happens
if we change the $x$-axis to be at 32$^o$ with respect to the horizontal direction. This is shown in Fig.~\ref{squeez2}.
We see that a relatively small change in the chosen strain protocol, in this case in $1^o$ in the chosen direction of the principal stress axes, results in a huge change in the position of the first plastic event. The aim of the rest of this Letter is to explain that this sensitivity
increases indefinitely with the system size, such that for macroscopic systems, i.e., in the
thermodynamic limit, any arbitrarily small change in protocol will result in a macroscopic
change in the position of the first plastic event.

\begin{figure}
\includegraphics[scale = 0.40]{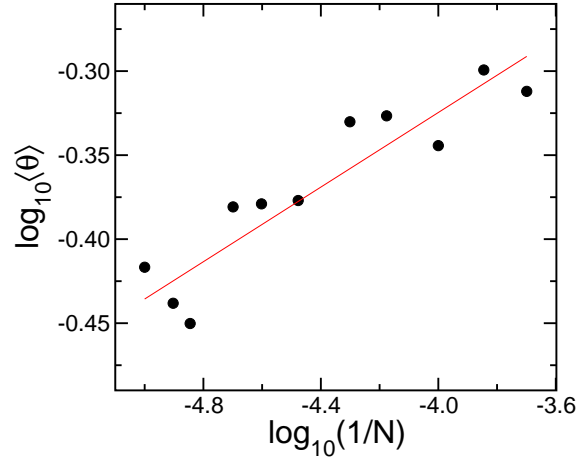}
\caption{The average angle required to observe a major change in the position of the first
plastic event as a function of the system size. Note the logarithmic scale used that
supports the power-law dependence Eq. (\ref{pl}). The systems studied here spanned the sizes $N=5,000$ to $N=100,000$,
all quenched at rate of $10^{-1}$.}
\label{thetN}
\end{figure}

To this aim we prepare between 30 to 100 different realizations of our system for each system size, changing the number of
particles in the range $N=5,000-100,000$. Each realization is then strained as described above,
choosing (arbitrarily) the $x$-axis to coincide with the original $x$-axis of the square box. For each realization we determine what is the first plastic event and what is the value of $\gamma_P$ where it appears. In a second step of this exercise we change the $x$-axis to have an angle $\theta$ with respect to the original horizontal direction. We then determine, for
each realization, the first angle $\theta$ for which {\em the first plastic event is different},
as seen in Figs. \ref{squeez1} and \ref{squeez2}. Finally, we average the angle $\theta$ over
the 100 realization to get $\langle \theta \rangle$ as a function of the system size $N$.
The central result of this exercise is that $\langle \theta \rangle(N)$ is a
decreasing function of $N$ as seen in Fig. \ref{thetN}.
A fit to the numerical data shown in Fig.~\ref{thetN} supports a power law of the form
\begin{equation}
\langle \theta\rangle(N)\sim N^{-\alpha} \ , \quad \alpha\approx 0.11\pm  0.02\ .
\label{pl}
\end{equation}
Clearly, in the thermodynamic limit $N\to \infty$, this result strongly indicates
that indeed any infinitesimal change in angle should result in a macroscopic change
in the position of the plastic event. This evidently refutes any possibility to predict
the position of the plastic event from the analysis of the system's state in equilibrium,
before straining. We should note that the similar data top those shown in Fig. \ref{thetN} were also
obtained with other quench rates with identical conclusions.

In summary, we have presented very simple tests to decide between two deeply contrasting views of the nature
of plastic events in amorphous solids. The evidence provided above indicates that in the thermodynamic limit
it is impossible to predict where the first plastic event should appear in a stressed amorphous solid. The plastic events
are protocol dependent, and any minute change in the protocol should result in a macroscopic change in the position of
the first plastic event. We conclude that it would be futile to predict the position of the first plastic event from
analyzing the structure of the amorphous solid at equilibrium, be the method of analysis as sophisticated as one might
think of. It is important to stress at this point that our analysis also indicate that later plastic events are even
more sensitive to the change in protocol, and the system size dependence of their sensitivity is more
steep than the findings reported in Eq. (\ref{pl}). This and related findings are however beyond the scope of this
Letter which aims specifically to sharpen the difference in the current approaches to plasticity in amorphous solids.
\vskip 0.5 cm 
\noindent {\bf Acknowledgments}\\
PKJ is supported by a PBC outstanding postdoctoral fellowship from the Council of Higher Education (Israel) for researchers from India. This work was supported by an ``ideas" STANZAS grant from the ERC.

\end{document}